\documentclass[aip, reprint, amsmath,amssymb,apl]{revtex4-1}
\draft 
\usepackage{graphicx}
\usepackage{mathptmx}
\usepackage{subcaption}
\usepackage{hyperref}
\begin{document}


\title{Experimental Demonstration of A Dual-Input/Dual-Output Reflective Impedance Metasurface} 

\author{Jean~Louis~Keyrouz}
\email[Author to whom correspondence must be addressed:]{jeanlouis.keyrouz@mail.utoronto.ca}
\author{Vasileios G. Ataloglou}
\author{George~V.~Eleftheriades}
\affiliation{The Edward S. Rogers Sr. Department of Electrical and Computer Engineering, University of Toronto, Toronto, Ontario M5S 3G4,
Canada}




\date{\today}

\begin{abstract}
This paper presents the experimental demonstration of a dual-input/dual-output reflective impedance metasurface. The design of the metasurface relies on the Method of Moments and leverages auxiliary surface waves to achieve anomalous reflection of two impinging plane waves with controlled sidelobe levels. The two beams are chosen independently compared to those in a conventional phase-gradient metasurface where the design presents a single slope to achieve a certain reflection and all other incident beams would depend on that slope. A prototype that ensures maximum directivity at two prescribed reflection angles for the two input waves is then fabricated on a Rogers RO3003 printed-circuit board using 42 metawires loaded with printed capacitors. The proposed metasurface is capable of reflecting an incident beam from $-20^\circ$ to $-55^\circ$ and a second from $+10^\circ$ to $50^\circ$ at 9.93 GHz. The metasurface is experimentally characterized and an illumination efficiency of at least 89\% is calculated for each of the reflected waves, indicating a high multiplexing efficacy.
\end{abstract}


\maketitle 

The advent of new wireless communication systems has elicited a renewed interest in the development of energy efficient and compact beamforming antennas using metasurfaces (MTSs). In fact, MTSs have been used in a wide range of applications showcasing their exceptional ability to mold electromagnetic waves in various ways, such as perfect anomalous refraction \cite{Selvanayagam,Pfiefer}, perfect anomalous reflection \cite{Wong,Alu}, beamforming \cite{Dorrah,Roeker}, and polarization control \cite{Selvanayagam2,Niemi}. 

Metasurfaces have also facilitated the design of structures that radiate multiple independent beams, when fed by multiple spatially-separated feed points. Each feed point produces a beam in a different direction, enabling multiple-input multiple-output (MIMO) communications with the use of a single aperture. In the context of holographic antennas, the design with multiple feeds and multiple output beams is typically done by averaging the impedance of all the different cases \cite{Vas1,Maci,Lemberg}. The same method of multiplexing has also been proposed in the case that the feeds operate at different frequencies \cite{Vas2,Vas3}. While the averaging multiplexing method provides some enhancement in the aperture efficiency of all beams compared to partitioning the metasurface into different regions, it still suffers from cross interference-terms, namely the radiation that is produced when the wave from each feed interacts with the modulation designed for the other feeds. Therefore, unless there is orthogonality of the incident fields/modulations or the positions of the feeds is carefully optimized, the total performance would be sub-optimal. A solution was proposed in \cite{Vas4} by forming an over-defined stacked system of equations for the unknown impedances based on integral equations and the desired radiated fields. This is solved in a quasi-direct approach using the least square method \cite{Bod}.

On the other hand, integral equation models solved using the Method of Moments and using a classic iterative optimization approach for determining the necessary impedances for desired functionalities have been recently proposed \cite{Bhudu1, Xu:Access2022}. While these methods directly optimize for the far-field pattern, the underlying mechanism is the excitation of auxiliary surface waves that redistribute the power passively along the MTS \cite{Epstein,Kwon2018,Vasilis}. In addition to capturing the mutual coupling interactions among the different unit cells comprising the MTS, integral equation methods enable the realization of MIMO scenarios. The iterative optimization approach has certain benefits as the exact solution for all inputs is available at each iteration during optimization. In the case of multiple incident fields, this means that the optimization scheme treats all cases rigorously, and there are not any unpredicted cross-interference terms. Moreover, constraints can be set on the range of the impedance values so that the latter are physically realizable. Lastly, a set of metrics such as the directivity or the sidelobe level can be introduced in the cost function to balance the requirements of each application. Using this approach, a multi-layered structure has been proposed for shaping the reflected radiation pattern for two different frequencies \cite{Bhudu2}.

In this paper, we discuss the design and measurement of a MIMO single-layer reflective impedance MTS that anomalously reflects two independent plane waves in an asymmetric manner. This is in contrast to phase-gradient metasurfaces that can be designed to anomalously reflect an incident wave in a desired direction but with no control over the reflection of other incident waves\cite{light}. Moreover, by breaking the periodicity, the possible output angles are not constrained by Floquet-Bloch modes as in coarsely-discretized metasurfaces (often, referred to metagratings)\cite{Wong2}. The efficient anomalous reflection of two incident beams effectively doubles the number of channels that the MTS can handle simultaneously. Such reflective passive MTSs can be used to redirect multiple beams independently in an indoor or outdoor wireless environment\cite{Saluzzi}. 

The metasurface is designed based on the integral-equation optimization framework to maximize the directivity of two reflected plane waves at predetermined angles, while keeping the sidelobes at a tolerable level \cite{Xu:Access2022}. While the design is based on the same optimization method, this paper addresses some practical issues compared to the MIMO example presented therein. Specifically, the simulation is not based on idealized impedance sheets, but on realistic strip wires that are loaded with printed capacitors adhering to all fabrication tolerances. In turn, this allows the estimation of losses and bandwidth of the proposed design, which are important for any metasurface used for communications applications. It is noted that losses can be significant if high surface waves are excited, as in the case of the unconstrained solution presented before \cite{Xu:Access2022}. Finally, the design is fabricated on a Rogers RO3003 printed-circuit-board with sub-wavelength printed capacitor unit cells etched on its surface resulting in a compact and lightweight structure. Compared to the theoretical model, the prototype is finite in all dimensions, adding to the non-idealities of the experiment. Nevertheless, the reflections of two impinging planes waves are measured in a bi-static setup at $9.93$ GHz, and they match satisfactorily with theory, further validating the design. The MoM framework used to design MIMO MTSs allows the reflection of independent angle pairs that we detail in the supplementary material.

\begin{figure}
\centering
\includegraphics[width=0.5\columnwidth]{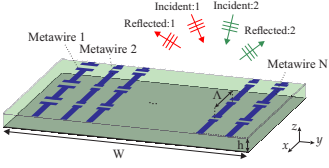}
\caption{Model of the impedance MTS with printed capacitor unit cells. }
\label{fig:TheoMTS}
\end{figure}

The designed structure consists of a set of metawires on top of a grounded dielectric substrate of thickness $h$ and relative permittivity $\varepsilon_r$. The substrate has a width of $W$ along the y direction while the metawires are assumed to be infinite along the x direction with periodic capacitive loadings every $\Lambda$. The very sub-wavelength nature of the periodic loadings allows us to assign a homogenized loading impedance $Z_i$ at every metawire (i). The different impedance loadings $Z_i$ are implemented using printed capacitors of varying lengths, as shown in Fig. \ref{fig:TheoMTS}. It is noted that although the MTS in Fig.\ref{fig:TheoMTS} is intended for operation in free-space, nearby metallic or dielectric scatterers could have been included and accounted for rigorously, as it has been done in the context of reconfigurable intelligent surfaces \cite{rev2}.

The MTS is illuminated with transverse electric (TE) plane waves ($\mathbf{E}_\mathrm{inc}^n=E_\mathrm{inc}^n \mathbf{\hat{x}}$) at specific incident angles $\theta_{in}$ in the yz plane with positive angles extending in the +y direction and negative angles extending in the -y direction, with $n$ representing the different inputs.
Each incident field $E_{\mathrm{inc}}^n$ induces surface conduction currents $J_g$ and $J_w$ in the ground plane and the metawires respectively, as well as, volume polarization currents $J_v$ in the dielectric. The scattered fields generated by these three induced currents, in addition to the incident field, form the total field radiated by the MTS. A set of coupled equations for $J_g$,$J_v$ and $J_w$ is formed by inspecting the total electric field at the different regions. The total electric field must be zero on the ground plane because it is entirely tangential to it. According to Ohm's law, the total electric field is proportional to $J_w$ with a proportionality constant of $Z_i$ at every metawire (i) and proportional to $J_v$ in the dielectric substrate. Expanding the unknown conduction and polarization currents using the MoM into pulse basis functions and using point matching at the center of each basis function leads to the following linear system of equations \cite{Xu:Access2022}: \begin{align}\label{eq:MoM_system}
\left(
\begin{bmatrix}
0 & 0 & 0\\
0 & P & 0\\
0 & 0 & Z_w
\end{bmatrix}-
\begin{bmatrix}
G_{gg} & G_{gv} & G_{gw}\\
G_{vg} & G_{vv} & G_{vw}\\
G_{wg} & G_{wv} & G_{ww}
\end{bmatrix}\right)
\begin{bmatrix} 
\bar{J_g} \\
\bar{J_v} \\
\bar{J_w}
\end{bmatrix}=\begin{bmatrix} 
\bar{E_i^g} \\
\bar{E_i^v} \\
\bar{E_i^w}
\end{bmatrix},
\end{align}
where $P$ and $Z_w$ are diagonal matrices related to the electric susceptibility of the substrate and the loading impedances metawires $Z_i$ respectively, the $\bar{J}$ entries represent the sampled induced currents, the G entries represent the self and mutual interactions between the various components of the impedance MTS and the $\bar{E}_{i}$ entries represent each incident field sampled at the ground plane, substrate and metawires. For a given set of loading impedances $Z_i$ (encapsulated in the matrix $Z_w$), Eq.~\eqref{eq:MoM_system}  can be used to solve the forward problem; namely, to find the induced currents and then calculate the scattered fields in the far-field region. A more detailed description of the integral equation framework used in our analysis can be found in previous publications \cite{Xu:Access2022}.

Having established a way to calculate the scattering for a given set of impedance loadings $Z_i$, the latter can be optimized to match  a desired radiation pattern for any given incident field $E_\mathrm{inc}^n$, since Eq.~\eqref{eq:MoM_system} is valid for any illumination of the MTS. Naturally, different incident fields in Eq.~\eqref{eq:MoM_system} would result in a different set of induced currents and, thus, different far-field patterns for each input. In fact, if we consider a dual-input/dual-output system (n=1,2), the goal is to optimize for a single set of impedances $Z_i$ that will allow both incident electric fields to achieve the desired far-field radiation patterns $D^{ff}_n$. It should be pointed out that all metawires contribute to both functionalities at a collective manner, and there are not some wires designated for the one or the other case. This feature makes possible the effective utilization of the entire aperture for both cases, simultaneously.

In this work, we primarily aim to achieve maximum directivity at the desired output angles $\theta_{on}$, while maintaining a satisfactory sidelobe level. The cost function that quantifies these features for two incident angles (n=1,2) can be defined as follows:
\begin{equation}
\label{eq:costfunc}
F=-\sum_{n=1}^2 D^{ff}_n(\theta_{on})+\sum_{n=1}^2 
\mathrm{max}\{D^{ff}_n (\theta_{on}^\mathrm{SLL})-S_n,0\},
\end{equation} where $D^{ff}_n(\theta_{on})$ represents the directivity at the desired output angles $\theta_{on}$, $D^{ff}_n (\theta^{\mathrm{SLL}}_{on})$ represents the directivity at a set of angles $\theta^{\mathrm{SLL}}_{on}$ that excludes the main beam, and $S_n$ represents the maximum level the sidelobes can reach, all in dB scale. We note that the cost function is minimized with respect to the impedance loadings without going through a calculation of a local reflection coefficient for each wire. Specifically, the imaginary parts of the impedance loadings of the wires are the ones being optimized, since they play a crucial role in the emerging pattern. On the other hand, the real part corresponds to the small unavoidable losses of a physical implementation and it only limits the power efficiency of the metasurface. Due to the physical limitations when realizing the sub-wavelength unit cells, a constraint is placed beforehand on the range of values for the imaginary part of the impedance loadings $\mathrm{Im}\{Z_i\}$. 
The first term in Eq.~\eqref{eq:costfunc}  guarantees maximal directivity in the chosen directions $\theta_{on}$ while the second term prevents any sidelobe defined in the set of angles $\theta^{\mathrm{SLL}}_{on}$ from rising above a predetermined absolute level $S_n$. The values $S_n$ are determined by subtracting the desired sidelobe level from the expected maximum directivity for each beam.
Minimization of the cost function is done in MATLAB using the built-in genetic algorithm followed by gradient descent optimization. The output of the genetic algorithm is inherently random, so multiple iterations of the optimization algorithm can be performed before selecting the solution with the optimal pattern. Finally, it is noted that while the beams and the specifications are equally weighted in forming the cost function of Eq.~\eqref{eq:costfunc}, unequal multiplicative factors could have been included if one specification (directivity or sidelobe level) or one beam was considered of higher importance.

\begin{figure}
\centering
\includegraphics[width=0.5\columnwidth]{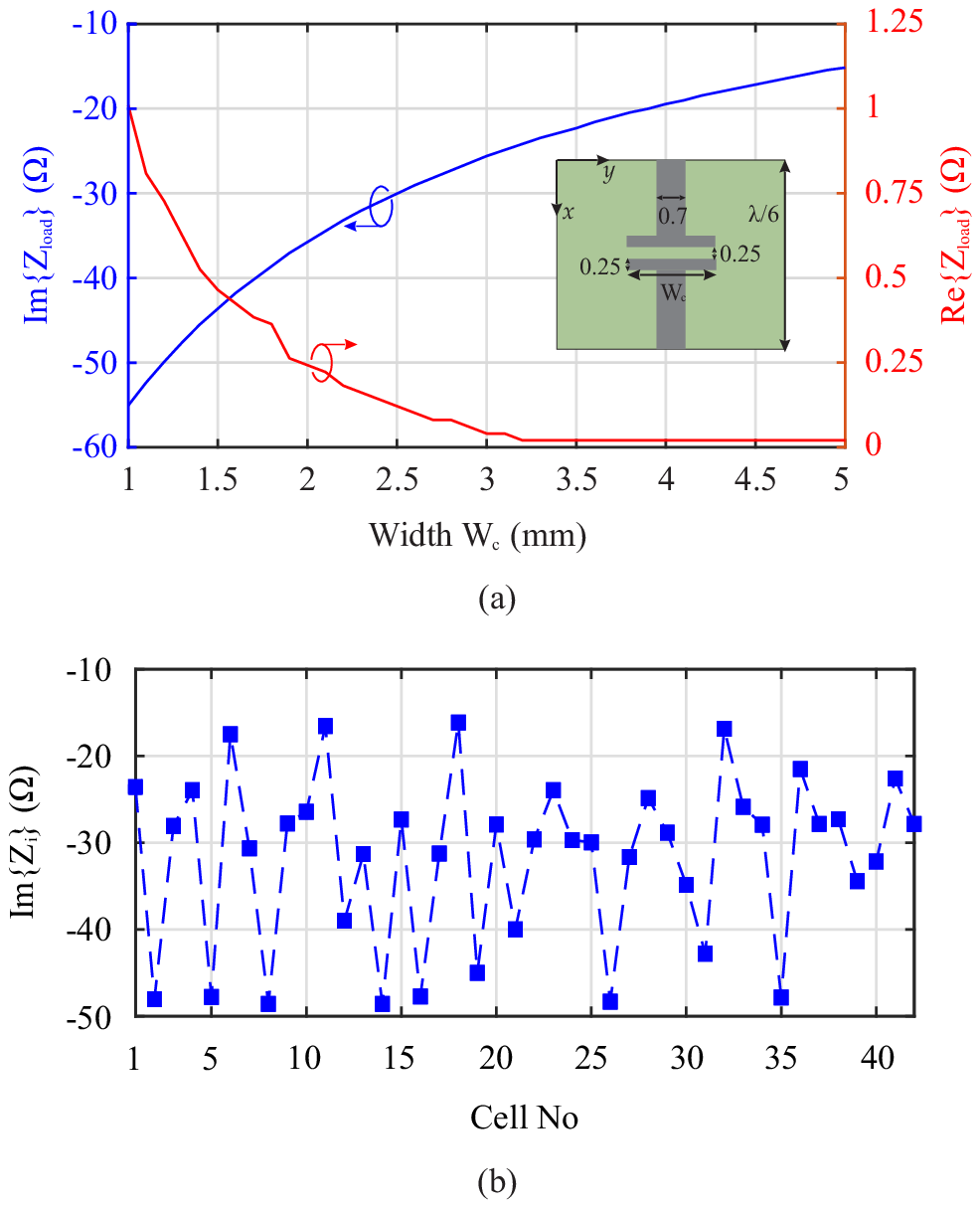}
\caption{(a) Impedance loading variation $Z_\mathrm{load}$ as a function of the length of a printed capacitor.
(b) Optimized set of loading impedances $\mathrm{Im} \{Z_i\}$ for the 42 metawires.}
\label{fig:ImpedanceVariation}
\end{figure}

For the presented example, only capacitive loadings are utilized, since the achievable range is sufficient to realize the desired functionality. The loadings are implemented as printed capacitors that give a reliable and cost-efficient method to realize the required impedances. A correspondence between the capacitive loading $Z_\mathrm{load}$ and a printed capacitor is established by characterizing a single capacitive unit cell in Ansys High Frequency Structure Simulator (HFSS). In fact, an embedded source is placed below the unit cell and a full-wave simulation is performed to record the electric field emanating from the structure for different capacitor lengths. A similar procedure is then performed with a homogenized strip of a varying impedance $Z_\mathrm{load}$ using the MoM framework. By comparing the scattered near-field from the HFSS simulation (involving the patterned wires) and the MoM model (involving the impedance sheets), a correspondence between the capacitor length and the associated impedance loading can be established, as shown in Fig. \ref{fig:ImpedanceVariation}(a). Using the above-described procedure to characterize a single metawire avoids placing it in an infinite periodic array, typically found in metasurface or reflectarray designs. This is aligned with the MoM optimization framework which models the homogenized impedance loading of a single metawire and includes all mutual coupling interactions between dissimilar neighboring metawires.

The design and measurement of a reflective impedance MTS that displays the anomalous reflection of two incident plane waves at desired angles are outlined.
The designed MTS consists of 42 metawires with a loading period of $\Lambda=\lambda/6$ on top of a grounded substrate of height $h=1.52 \ \mathrm{mm}$, permittivity $\varepsilon_r=3$ and loss tangent $\mathrm{tan} (\delta)=0.001$. It has a width of W = $8 \lambda$ at 10 GHz. The MTS reflects two incident beams $35^\circ$ and $40^\circ$ off their specular direction. In fact, two uniform planes waves impinging on the surface at $\theta_{i1}=-20^\circ$ and $\theta_{i2}=+10^\circ$ are anomalously reflected at $\theta_{o1}=-55^\circ$ and $\theta_{o2}=+50^\circ$ respectively. These two pairs of angles can be chosen independently of each other with the design satisfying both reflections simultaneously.

The optimization algorithm previously discussed was run multiple times to obtain a pattern with a sufficiently low cost function value. The sidelobe level parameters in Eq.\eqref{eq:costfunc} $S_1$ and $S_2$ were set in a way that ensures a sidelobe level at most equal to $-13.3 \ \mathrm{dB}$ when measured from the peak to maintain a high efficiency. The optimization is performed on the imaginary parts of the loading impedances $\mathrm{Im}\{Z_i\}$, as the effect of the real parts $\mathrm{Re}\{Z_i\}$ on the radiation pattern is minimal. The range of $\mathrm{Im}\{Z_i\}$ was constrained to $[-50, -15] \Omega$ and the optimized loading impedances $Z_i$ are shown in Fig. \ref{fig:ImpedanceVariation}(b). 

\begin{figure}
\centering
\includegraphics[width=0.5\columnwidth]{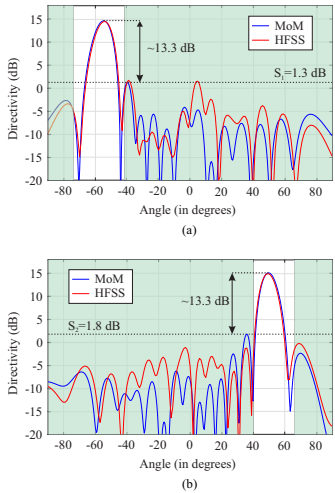}
\caption{MoM and HFSS directivity plots for (a) $-20^\circ$ to $-55^\circ$ and (b) $+10^\circ$ to $+50^\circ$ at 10 GHz. The green shaded regions represent the ranges $\theta_{on}^{SLL}$ that exclude the main beam, and $S_n$ represent the constraint levels imposed on Eq.\eqref{eq:costfunc} for the optimization.
}
\label{fwhat}
\end{figure}

The results from the MoM model and the HFSS simulation involving the physical structure are shown for each case in Fig. \ref{fwhat}. They both present a high directivity and satisfactory sidelobe levels of at most -12.8 dB and similar patterns at 10 GHz.
We calculate the ideal 2D directivity of an aperture with uniform-amplitude based on \cite{GlebTAP}: 
\begin{equation}
D^{ff}_{uni}(\theta_{on}) = \frac{2 \pi W}{\lambda}\cos(\theta_{on}).
\end{equation} 
The interpretation of 2D directivity in Eq. (3) is discussed further in the Supplementary material.
The illumination and power (radiation) efficiencies, which are defined as,
\begin{align}
\eta_{il} = \frac{max(D^{ff}(\theta))}{D^{ff}_{uni}(\theta_{on})}      && 
\eta_{p} = \frac{P_{ref}}{P_{inc}}
\end{align}
are used to calculate the total efficiency of the HFSS radiation pattern where $D^{ff}(\theta)$ represents the directivity pattern, $P_{ref}$ the reflected power and $P_{inc}$ the incident power. It is highlighted that the illumination efficiency relates to the directivity while the power efficiency is related to losses. For the wave incident at $\theta_{i1}=-20^\circ$ and reflected at $\theta_{o1}=-55^\circ$, the simulated efficiencies are $\eta_{il} = 0.995$ and $\eta_{p} = 0.906$; therefore, the total efficiency is $\eta = \eta_{il}\eta_{p} = 90.1\%$. Similarly, for the wave incident at $\theta_{i2}=+10^\circ$ and reflected at $\theta_{o2}=+50^\circ$, $\eta_{il} = 0.991$, $\eta_{p} = 0.886$ and the total efficiency is $\eta = \eta_{il}\eta_{p} = 87.8\%$.
The simulated fractional bandwidth for a 3-dB drop in the gain is 4.16\% and 8.95\% for the reflected beams directed at $-55^\circ$ and $+50^\circ$, respectively. The bandwidth is mostly limited by the frequency dispersion of the capacitive impedance loadings, as well as by the change of all the electric distances as the frequency varies. 

The length of the MTS in the wire longitudinal direction (x-axis) was truncated to around 6$\lambda$. Using finite-length homogenized impedance sheets, full-wave simulations in HFSS showed that such a truncation is sufficient to approximate the assumed uniformity along the x-axis and maintain the far-field patterns shown in Fig.\ref{fwhat}. The design was implemented on a 17.99 cm by 23.98 cm board using a Rogers RO3003 substrate with the metawires etched on one side while the other side is kept copper laminated to serve as a ground plane.

\begin{figure}
\centering
\includegraphics[width=0.5\columnwidth]{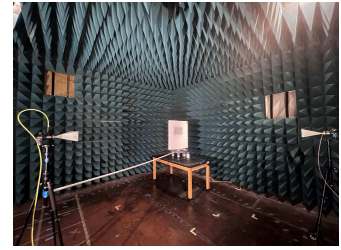}
\caption{Bi-static measurement setup. The transmitting antenna illuminates the MTS while the rotating receiving antenna records the reflected pattern.
}
\label{f4}
\end{figure}

The antenna is measured in an anechoic chamber with two identical X-band horn antennas acting as the transmitter and receiver, as shown in Fig. \ref{f4}. The apertures of the horn antennas are around 245 cm away from the MTS. The transmitting horn antenna is fixed to the ground while the MTS is rotated by $20^\circ$ and $10^\circ$ to emulate an incidence at $-20^\circ$ and $+10^\circ$, respectively. Measurements of $S_{21}$ are performed with a resolution of $1^\circ$ near the main beam and up to $5^\circ$ for the rest of the pattern. The horn antennas occupy a region of around $7.5^\circ$ each, which prohibits measurements of the reflected beam in a $15^\circ$ region around the incident angle. Both the receiving and transmitting horn antennas are mounted on tripods connected by plastic tubes to the platform on which the MTS is placed. This ensures the receiving antenna is maintaining a constant distance from the MTS when measurements are taken at the different angles.

\begin{figure}[h]
\includegraphics[width=0.5\columnwidth]{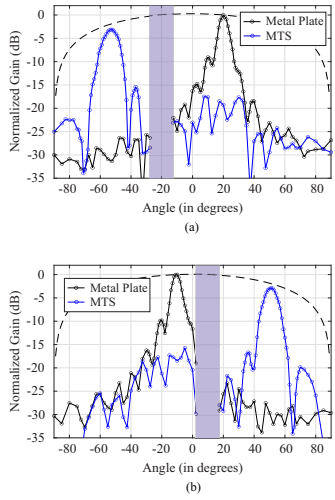}
\caption{Normalized Gain radiation patterns of the measurements and copper plate at 9.93 GHz. The shaded areas represent the zones where no measurements are possible ($[-27.5^\circ,-12.5^\circ]$ for (a) the first incident beam and (b) $[2.5^\circ,17.5^\circ]$ for the second incident beam).}
\label{f6}
\end{figure}

The measurement results are shown in Fig. \ref{f6} alongside the measurements performed at the same incident angles on a copper plate of similar dimensions, both normalized to the maximum directivity of the copper plate at 9.93 GHz, that maximized the MTS gain in the desired angles. The measured gain pattern is similar to the pattern simulated in HFSS. The specular direction gain is heavily suppressed demonstrating the efficacy of the MTS in reflecting the input beams at the desired directions. Both patterns present sidelobe levels of at least 12.3 dB below the peak directivity. The maximum gain of the first measured pattern occurs at $\theta_{o1}=-53^\circ$ and it is around 1.19 dB (76\% efficiency) lower than the maximum expected gain if the copper plate redirected the incident power in the desired direction. This maximum is represented by the dashed line in Fig. \ref{f6}. Similarly, the maximum gain of the second measured pattern occurs at $\theta_{o2}=51^\circ$ with a gain 0.96 dB (80\% efficiency) lower than the gain of the copper plate. The directivity of the pattern is calculated using the measurement results. The illumination and subsequently the power efficiency of the two beams is then calculated. For the first measured pattern, the illumination efficiency is 89\% and using the measured total efficiency (76\%), we calculate a power efficiency of 85\%. Similarly, for the second measured pattern we calculate an illumination efficiency of 93\% and a power efficiency of 86\%. A different way of estimating the total efficiency and the 3-dB bandwidth for each beam through the bi-static radar equation is discussed in Sec. III of the Supplementary material.

In conclusion, we presented an experimental demonstration of a reflective impedance metasurface that handles two independent incident beams simultaneously. The uniform plane waves impinging on the surface at $-20^\circ$ and $+10^\circ$ are anomalously reflected to $-55^\circ$ and $+50^\circ$ at $9.93$ GHz, respectively, while keeping the sidelobe level at most -12.3 dB . The metasurface was designed using a Method-of-Moments based optimization framework and simulated with printed capacitor unit cells which make the analysis of losses and bandwidth possible. The physical design was fabricated and measured for both input channels reporting highly directive beams in the desired directions, as evident from the radiation pattern and the high total efficiencies. 

%
%

%


\section*{Data Availability}
The data that support the findings of this study are available from the corresponding author upon reasonable request. Additional information concerning the experiment can be found in the Supplementary information.

\section*{Supplementary Material}
See the supplementary material for details on directivity, bandwidth and efficiency calculations, as well as the versatility of the framework compared to phase-gradient metasurfaces and its ability to support more beams.
\begin{acknowledgments}
This work was supported by the Department of National Defence’s Innovation
for Defence Excellence and Security (IDEaS) Program.
\end{acknowledgments}

\end{document}